\documentclass[11pt, a4paper]{article}

\usepackage[utf8]{inputenc}
\usepackage[T1]{fontenc}
\usepackage{mathptmx}             % Times New Roman style text
\usepackage[margin=1in]{geometry} % Standard margins
\usepackage{microtype}            % Improves spacing and kerning (prevents bad line breaks)
\usepackage{titlesec}
\usepackage{authblk}
\usepackage{enumitem}
\usepackage{booktabs}
\usepackage{graphicx}
\usepackage{url}

\usepackage[colorlinks=true,
            linkcolor=blue,
            citecolor=blue,
            urlcolor=blue,
            pdftitle={Representative Litigation Settlement Agreements in AI Copyright Disputes},
            pdfauthor={Chanhou Lou}]{hyperref}

\title{\textbf{Representative Litigation Settlement Agreements in Artificial Intelligence Copyright Infringement Disputes: A Comparative Reflection Based on the U.S. \textit{Bartz} Case}}

\author{Chanhou Lou}
\affil{\small University of Macau; Cornell University}
\date{\today}

\begin{document}

\maketitle

% --- Abstract ---
\begin{abstract}
\noindent The high-density, decentralized copyright conflicts triggered by generative AI training require more than ad hoc solutions; they demand structural governance tools. This article argues that \textit{representative litigation settlement agreements} offer a distinct institutional advantage. Beyond reducing the transaction costs associated with the ``tragedy of the anticommons,'' these agreements generate market-visible evidence, specifically pricing signals and licensing practices, that validate the ``potential market'' under the fourth factor of fair use. This phenomenon constitutes \textit{procedural market-making}. Through a comparative analysis of the U.S. \textit{Bartz} class action settlement, this study reveals a dual motivation: a surface-level drive for risk aversion and remedy locking, and a deeper logic of constructing a \textit{training-licensing market}. In the context of Chinese law, the feasibility of such agreements depends not on replicating foreign models, but on establishing three interpretive mechanisms: expanding the functional definition of ``same category'' claims; adopting a hybrid registration/confirmation system for \textit{indeterminate class membership}; and converting the ``consent'' requirement under Article 57, Paragraph 3 of the Civil Procedure Law into a workable \textit{opt-out} right subject to judicial scrutiny.

\vspace{1em}
\noindent \textbf{Keywords:} Generative AI; Representative Litigation; Class Action; Settlement Agreement; Copyright Disputes; Fair Use; Training-Licensing Market.
\end{abstract}

\vspace{1em}
\hrule
\vspace{1em}

% --- Main Content ---

\section{Introduction}

The copyright infringement disputes triggered by generative AI models (hereinafter ``AI copyright disputes'') have garnered significant academic attention. However, existing scholarship predominantly focuses on substantive law issues, such as whether infringement is negated by the fair use defense \cite{Wan2019, WangW2022, Wu2025, Wu2020}, while largely neglecting procedural questions regarding dispute resolution through settlement agreements \cite{Xiao2024}. This ``substance-over-procedure'' bias fails to recognize that procedural settlement agreements can exert a profound factual influence on substantive fair use determinations.

Taking the U.S. case \textit{Bartz et al. v. Anthropic PBC} (hereinafter ``\textit{Bartz}'') as a comparative example \cite{Bartz2025a}, the defendant AI platform and the plaintiff copyright holders reached a class action settlement agreement totaling \$1.5 billion \cite{Bartz2024}. This article posits a theoretical hypothesis worth verifying: settlement agreements regarding training works transcend the individual case to create factual precedents for similar future cases. They constitute evidence of a ``training market'' for copyright works, thereby supporting copyright holders in asserting the ``market effect'' theory to defeat fair use defenses.

This logic is defined here as ``\textit{procedural market-making}.'' The validation of a copyright licensing market for AI training helps plaintiffs argue, under the fourth factor of the U.S. copyright fair use test \cite{CopyrightAct}, that the unauthorized use produces a market substitution effect. Consequently, the incentive for copyright holders to join the \textit{Bartz} settlement extends beyond the ``scale effect'' of reducing individual costs; it includes the strategic benefit of using collective settlement to define the \textit{training-licensing market} as a \textit{copyright-adjacent market}, thereby securing future copyright revenues.

Compared to the \textit{Bartz} case, the use of representative litigation settlements in Chinese copyright practice is rare. Although China's \textit{Civil Procedure Law} provides for representative litigation and court-facilitated mediation \cite{CPL}, these tools are seldom utilized for copyright disputes. Furthermore, while the \textit{Provisions on Representative Litigation in Securities Disputes} (2020) introduced mechanisms similar to U.S. class actions, they are restricted to the securities sector \cite{SPC2020a}.

Therefore, this article addresses two questions: First, why are \textit{representative litigation settlement agreements} appropriate for resolving AI copyright disputes? Second, if appropriate, how can their feasibility be established within the Chinese legal context?

Methodologically, this article does not treat \textit{Bartz} as an ideal model to be copied, but as a field study in critical reflection. Drawing on materials from the \textit{Bartz} settlement roadshow \cite{FieldNotes2026} and analysis of related procedural documents \cite{AnthropicWeb}, this study identifies the empirical challenges in U.S. practice to generate comparative insights for China.

\section{The Appropriateness of Class Action Settlements: The ``Tragedy of the Anticommons'' and ``Dual Motivations''}

An analysis of \textit{Bartz} demonstrates the appropriateness of using collective mechanisms for AI disputes. The case, presided over by Judge William Alsup in the N.D. California, centered on whether Anthropic's use of books from \textit{shadow libraries} (e.g., LibGen) for \textit{large language model (LLM)} training constituted fair use \cite{Bartz2025b}.

Procedurally, the court's summary judgment ruling distinguished between ``model training'' and ``pirated downloading.'' While the court inclined toward fair use for training, it found that downloading from pirate sites to build a database was likely not fair use \cite{Bartz2025b}. Furthermore, the court issued a \textit{class certification} order covering all ``legal owners'' and ``beneficial owners'' (authors) whose works were in the datasets \cite{Bartz2025c}. This certification significantly increased the defendant's risk exposure, facilitating a \$1.5 billion settlement \cite{Bartz2025d}.

\subsection{The ``Tragedy of the Anticommons'' in AI Copyright Disputes}

Generative AI infringement is characterized by the use of datasets containing hundreds of thousands of works, creating a ``tragedy of the anticommons'' where excessive fragmentation prevents effective resource utilization \cite{Heller1998}. The dispersed nature of rights holders leads to prohibitive transaction costs.

Both sides in \textit{Bartz} utilized this difficulty to construct their arguments. The defendant argued that high transaction costs proved market failure, justifying fair use. Conversely, plaintiffs argued that individual copyright holders face ``\textit{rational apathy},'' where the cost of enforcement outweighs the potential statutory damages, necessitating a \textit{class action} to aggregate claims and reduce marginal costs \cite{Bolodeoku2007, Xiong2014}.

\subsection{Explicit Motivations in the Settlement Agreement}

The settlement documents reveal several explicit motivations driving the resolution:

\begin{enumerate}
    \item \textbf{Risk Avoidance:} Plaintiffs sought to avoid the risk of ``innocent infringement'' verdicts, which could lower statutory damages to \$200 per work, compared to the estimated \$3,000 settlement payout \cite{Bartz2025d}.
    \item \textbf{Remedy Locking:} The settlement includes \textit{dataset-destruction obligations}, creating an \textit{injunctive effect} that establishes strict compliance standards for ``data source legitimacy'' \cite{Lou2025, BartzSettlement}.
    \item \textbf{Representative Incentives:} Class representatives received service awards (up to \$50,000), and class counsel secured risk-adjusted fees (approx. 20\% of the fund), incentivizing the organization of dispersed claimants \cite{Rubenstein2025, BartzFees}.
    \item \textbf{Business Preservation:} The defendant avoided the risk of ``willful infringement'' damages (up to \$150,000 per work) and preserved its core AI models, as the settlement did not require model destruction \cite{Bartz2025e, BartzSettlement}.
\end{enumerate}

\subsection{The Implicit Motivation: Procedural Market-Making}

Beyond these explicit factors lies a ``potential motivation'': \textit{procedural market-making}. This concept explains how the settlement retroactively impacts the ``potential market'' analysis under Factor Four of the fair use test \cite{CopyrightAct}.

In the wake of \textit{Andy Warhol Foundation v. Goldsmith} (2023), the U.S. Supreme Court has emphasized Factor Four to curb the overuse of the ``transformativeness'' test under Factor One \cite{Harper}. Defendants typically argue that a \textit{training-licensing market} is not a relevant \textit{copyright-adjacent market} \cite{Pitofsky1990}. However, a high-value settlement serves as empirical evidence that such a market exists.

\textit{Procedural market-making} generates three legal effects:
\begin{enumerate}
    \item \textbf{Market Pricing:} The \$3,000-per-book settlement establishes a benchmark price, refuting the ``market failure'' defense.
    \item \textbf{Commercial Custom:} The settlement establishes a custom that AI companies must pay for training data, supported by industry associations like the Association of American Publishers (AAP) \cite{Bartz2025d}. This mirrors the development of the photocopy licensing market recognized in \textit{Texaco} (1994).
    \item \textbf{Fair Use Blocking:} The existence of a functioning market makes it difficult for defendants to claim their use causes no market harm, thereby blocking the fair use defense.
\end{enumerate}

\section{The Feasibility of Representative Litigation Settlement Agreements in the Chinese Legal Context}

Given the appropriateness of the \textit{Bartz} model, is it feasible in China? While Article 57 of the \textit{Civil Procedure Law} provides for \textit{representative litigation} with \textit{indeterminate class membership}, it is rarely used for copyright \cite{Hanhan, Guangzhou}. Feasibility requires overcoming specific interpretive hurdles.

\subsection{Interpretive Difficulty: ``Same Category'' of Claims}

Article 57 requires that the ``subject matter of the action is of the same category.'' Traditional theory interprets this narrowly as identical substantive rights, which hinders mass tort actions \cite{Chen2024}. In AI disputes, while works differ in value, the \textit{pirated downloading} and training behaviors are identical facts, and the legal questions (infringement vs. fair use) are identical.
To ensure feasibility, ``same category'' should be interpreted functionally to cover ``individual claims arising from the same infringement facts and legal issues.'' Since settlement agreements rely on party autonomy rather than the preclusive force of a judgment, a broader interpretation is permissible.

\subsection{Interpretive Difficulty: Registration for Indeterminate Membership}

China traditionally follows a ``registration-based \textit{opt-in}'' model, unlike the U.S. ``\textit{opt-out}'' model \cite{Xiao1999}. However, the 2020 \textit{Provisions on Representative Litigation in Securities Disputes} introduced a ``hybrid'' model (Special Representative Litigation) where investor protection institutions register claims, and investors are deemed to participate unless they explicitly \textit{opt-out} \cite{Jiao2023}.

The \textit{Bartz} settlement also utilizes a hybrid approach:
\begin{enumerate}
    \item \textbf{Membership:} Defined by \textit{opt-out} (all owners are bound unless they exclude themselves).
    \item \textbf{Distribution:} Requires \textit{opt-in} (claimants must file a claim form to receive money) \cite{BartzNotice, BartzPlan}.
\end{enumerate}

This distinction is crucial for China. Copyright lacks a centralized ``public'' registry like the securities market. Therefore, a hybrid mechanism is necessary: use technical means (ISBN matching, hash value comparison) to define the class, while requiring an \textit{opt-in} process for identity verification and payment distribution. The following question is about cost allocation: Who bears the cost of this complex identification?
\begin{itemize}
    \item \textbf{Copyright Collective Management Organizations (CMOs):} While they act as ``arbitrary procedural representatives'' \cite{Liu2010}, their scope is limited by the ``approval system'' and vertical industry divisions \cite{Regs2013}. They may not cover all rights holders.
    \item \textbf{Private Lawyers (Contingency Fees):} Although ``mass litigation'' is currently on the negative list for contingency fees in China \cite{MOJ2021}, the \textit{Copyright Law} allows for ``reasonable expenses'' (lawyer fees) to be shifted to the defendant \cite{SPC2020b}. If fees are structured as independent settlement items subject to court review—similar to the \textit{Feile Audio} securities case (2020)—they can functionally replicate the incentives of contingency fees without violating administrative regulations.
\end{itemize}

\subsection{Interpretive Difficulty: The ``Consent'' Requirement}

Article 57, Paragraph 3 of the \textit{Civil Procedure Law} states that a representative's settlement must be ``consented to'' by the represented parties. A strict interpretation requiring unanimous consent renders mass settlements impossible (e.g., the \textit{Wuyang Bond} case where individual settlements occurred because collective consent failed) \cite{Wang2021, Wuyang}.

However, the judicial interpretation for securities disputes converts ``consent'' into an \textit{opt-out} right: if a party does not expressly object or exit after notice, they are deemed to agree \cite{SPC2020a, Kangmei}.
China should adopt this ``Exit Right'' approach for copyright. By providing due process protections, including notice, objection hearings, and judicial fairness hearings, the ``consent'' requirement can be satisfied procedurally, avoiding the transaction costs of obtaining individual signatures.

\section{Conclusion}

In the high-density, decentralized landscape of AI copyright conflicts, \textit{representative litigation settlement agreements} are not merely stopgap measures but structural governance tools. They overcome the ``tragedy of the anticommons'' and align remedy efficiency with industry compliance.
Through comparative reflection on \textit{Bartz}, this article reveals that such settlements are driven by a dual logic: risk aversion and \textit{procedural market-making}.
For implementation in China, feasibility relies on three interpretive shifts: (1) expanding ``same category'' claims to include AI training disputes; (2) adopting a hybrid registration/confirmation system for \textit{indeterminate class membership}; and (3) transforming the statutory ``consent'' requirement into a due-process-protected workable \textit{opt-out} right.

% --- References (Strict Order of Appearance) ---

\end{document}